\documentclass[MSNbibl,nameyear,dvips]{arxstspdf}
\usepackage{flushend}
\usepackage{stfloats}
\usepackage{graphicx}
\volume{29}
\issue{1}
\pubyear{2014}
\firstpage{9}
\lastpage{17}
\doi{10.1214/13-STS428} 



\begin{document}
\begin{frontmatter}

\title{Estimation of HIV Burden through Bayesian Evidence Synthesis}
\runtitle{Bayesian Evidence Synthesis for HIV}

\begin{aug}
\author[a]{\fnms{Daniela}~\snm{De Angelis}\corref{}\ead[label=e1]{daniela.deangelis@mrc-bsu.cam.ac.uk}},
\author[b]{\fnms{Anne M.}~\snm{Presanis}},
\author[c]{\fnms{Stefano}~\snm{Conti}}
\and
\author[d]{\fnms{A. E.}~\snm{Ades}}
\address[a]{Daniela De Angelis is Programme Leader, MRC Biostatistics Unit,
Cambridge Institute of Public Health, University Forvie Site,
Robinson Way,
Cambridge CB2 0SR, United Kingdom \printead{e1}.}%
\address[b]{Anne Presanis is Senior Investigator Statistician, MRC Biostatistics Unit,
Cambridge Institute of Public Health, University Forvie Site,
Robinson Way,
Cambridge CB2 0SR, United Kingdom.}
\address[c]{Stefano Conti is Statistician, Public Health England, Colindale Avenue, London NW9 5EQ, United Kingdom.}
\address[d]{Anthony E. Ades is Professor, School of Social and Community Medicine, University of Bristol, Bristol, BS8 2PS, United Kingdom.}
\affiliation{Cambridge Institute of Public Health, Cambridge Institute of Public Health,
Public Health England, University of Bristol}
\runauthor{De Angelis, Presanis, Conti and Ades}

\end{aug}

%
\begin{abstract}
Planning, implementation and evaluation of public health policies
to control the {h}uman
{i}mmunodeficiency {v}irus (HIV)
epidemic require regular monitoring of disease burden. This
includes the proportion living with HIV, whether diagnosed or not,
and the rate of new infections in the general population and in
specific risk groups and regions. Estimation of these quantities
is not straightforward: data informing them directly are not
typically available, {but} a wealth of indirect
information from surveillance systems and ad hoc studies can
inform functions of these quantities. In this paper we show how
the estimation problem can be successfully solved through a
Bayesian evidence synthesis approach, relaxing the
focus on ``best available'' data to which classical methods are
typically restricted.
This more comprehensive and flexible use of evidence
has led to the adoption of our proposed approach as the official
method to estimate HIV
prevalence in the United Kingdom since 2005.
\end{abstract}

%

\end{frontmatter}

\section{Introduction}\label{sec_intro}

The HIV disease is associated with serious morbidity, high costs of
treatment and care, and, in developing countries, with significant
mortality and a high number of potential years of life lost
(\cite{UNAIDS2010}). Planning for care provision and for implementation
and evaluation of public health policies to reduce transmission
{relies crucially} on robust monitoring of disease
burden. This burden includes the proportion (prevalence) living with
HIV; the proportion of infections remaining undiagnosed; and the rate
at which new infections occur (incidence), in both the general
population and in specific groups at high risk of infection and in
different locations. To acquire robust evidence on these quantities is
not easy. The assessment of HIV prevalence is complicated by the
absence of symptoms for a long time after infection. Incidence is even
more difficult to measure, requiring, at least, longitudinal
follow{-}up of uninfected individuals, with all the
complications of cohort studies.

Devising appropriate methods for estimation of prevalence and
incidence has generated a rich literature in the last 30 years
(\cite{Brookmeyer2010}, \cite{PresanisThesis}). For HIV prevalence ``direct''
methods have been particularly popular amongst the medical community
(e.g., \cite{McGarrigleEtAl}, \cite{LyerlaEtAl2006} and references therein)
for their apparent transparency. The underlying idea is that the
general population, of size $N$, is subdivided into $G$
nonoverlapping groups at different risk of acquiring HIV. Estimates
of proportions $\rho_g$ of risk group $g$ ($g=1,\ldots, G$) in the
population are multiplied by estimates of the prevalence $\pi_g$ of
HIV to produce a point estimate of the number of infected individuals
$N \pi_g \rho_g$ in each group and in the population $N \sum_g
\rho_g\pi_g= N \sum_g (\rho_g\pi_g \delta_g + \rho_g\pi_g
(1-\delta_g))$. Here $\delta_g$ denotes the proportion of infected
individuals diagnosed in group $g${,} and $N \sum_g \rho_g\pi_g \delta_g$
and $N \sum_g \rho_g\pi_g (1-\delta_g)$ represent the number of
diagnosed and undiagnosed infections in group $g$,
respectively. Typically, at least in developed countries with
concentrated epidemics like the United Kingdom (UK), the number of
diagnosed infections is known from surveillance schemes, so the
problem is to estimate the number of undiagnosed infections. Provided
direct data that measure size and prevalence for each group are
available, these methods are very simple and, consequently,
appealing. However, \textit{direct} information on all parameters is not
typically available, whereas there is a wealth of \textit{indirect}
information, from a variety of sources, which may inform functions of
the parameters of interest. This additional indirect information is
generally discarded as difficult to incorporate in this simplistic
framework. As a result, on one hand, unverifiable assumptions and ad
hoc adjustments are made to compensate for the lack of information. On
the other hand, an inefficient use is made of the information that is
available, with consequent imprecise and biased results due to the
selective nature of the data used in the estimation. Finally, in the
``direct'' methods there is no explicit model formulation, so it is not
possible to quantify formally the uncertainty surrounding the
resulting estimates or to validate them.

The statistical challenge is then to provide an inferential approach
{capable of combining} \textit{direct} and \textit{indirect}
information from
multiple sources and appropriately {accounting} for any uncertainty in the
data and parameters. The Bayesian paradigm naturally offers the most
appropriate framework to address this challenge (see Section~\ref{sec_discuss}). Bayesian synthesis of evidence from different
studies, perhaps {even those} with different designs{,} is not new
(e.g., \cite{EddyEtAl1992}, \cite{DominiciEtAl1999}, \cite{AdesSutton2006}) and is
{attracting} increasing attention with applications in various fields
(e.g., \cite
{SpiegelhalterBest2003}, \cite{ClarkEtAl2010}, \cite{GovanEtAl2010}, \cite{BirrellEtAl2011}).

In this paper, we describe how such an approach has been successfully
adopted to estimate HIV prevalence and incidence in England and Wales
(E\&W) in the population aged 15--44. The remainder of the paper is
organised as follows: the concept of Bayesian evidence synthesis is
defined in Section~\ref{sec_mpes}; the model to estimate HIV
prevalence is presented in Section~\ref{sec_prev}; a joint model for
prevalence and incidence is described in Section~\ref{sec_joint}; and
Section~\ref{sec_discuss} {offers a concluding} discussion.

\section{Bayesian Evidence Synthesis} \label{sec_mpes}
\begin{figure}

\includegraphics{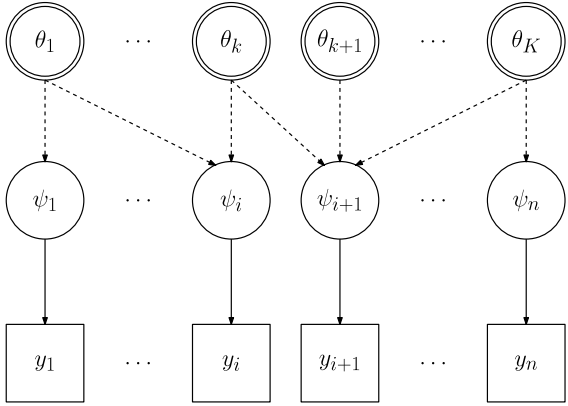}

\caption{DAG representation of a generic evidence
synthesis model.}
\label{fig DAG}
\end{figure}

Let $\bolds{\theta}=(\theta_1, \ldots, \theta_K)$ represent the
parameter vector we are interested in estimating. We refer to
$\bolds\theta$ as \textit{basic} parameters. Denote by $\mathbf
y=({\mathbf y}_1, \ldots,{\mathbf y}_n)$ a collection of $n\geq K$
independent data items available for the estimation of
$\bolds{\theta}$. Each ${\mathbf y}_i$ provides either \textit
{direct} information on a single component $\theta_k$ of
$\bolds\theta$ or \textit{indirect} information{, that is,} on
functional parameters, expressed in terms of one or more component(s)
of $\bolds\theta$. Denote by $\psi_i=\psi_i(\bolds\theta)$
a generic function of $\bolds\theta$, which may represent the
identity function{, that is,} $\psi_i=\theta_k$, a function of a single
parameter $\psi_i=\psi_i(\theta_k)$ or a function of multiple
components of $\bolds\theta$,
$\psi_i=\psi_i(\bolds\theta)$. Indicating by
$L_i(\psi_i(\bolds\theta); \mathbf y_i)$ the likelihood
contribution of ${\mathbf y}_i$ to the basic parameter vector
$\bolds{\theta}$, from the independence of the ${\mathbf y}_i$, the
full data likelihood is $L(\bolds\theta; \mathbf
y)=\prod_{i=1}^{n}L_i(\psi_i(\bolds\theta); \mathbf y_i)$. From a
Bayesian perspective, expressing the prior knowledge on
$\bolds{\theta}$ through a prior distribution
$p({\bolds\theta})$, inference is conducted on the basis of the
posterior distribution $p(\bolds\theta\mid\mathbf y)\propto
p({\bolds\theta}) L(\bolds\theta; \mathbf y)$, which
summarises all information, both \textit{direct} and \textit{indirect}, on
$\bolds{\theta}$. Such a distribution fully reflects the
uncertainty about $\bolds{\theta}$, including sampling
variability and parameter uncertainty, which automatically percolates
through to any function of the basic parameters $\bolds{\theta}$.
Figure~\ref{fig DAG} provides a direct acyclic graph (DAG)
(\cite{Lauritzen1996}) representation of the generic formulation above
and shows schematically the dependency between data and parameters as
well as the flow of information within the system. Here stochastic
``nodes'' are represented by circles and observed {``nodes''} by squares. The
basic parameters, in double circles, are given prior (possibly
hierarchical) distributions. Solid arrows represent distributional
assumptions{,} and dashed arrows indicate functional relationships. Note
the examples of functional parameters that inform multiple components
of $\bolds\theta$, such as
$\psi_i=\psi_i(\theta_1,\theta_k)$. Information flows along the
arrows, from the prior and from the data. The posterior distribution
of each $\theta_k$ is based on its prior distribution and on direct
and indirect information available on it, as well as the priors and
information on other components of $\bolds\theta$.

\begin{table*}
\caption{Relationship between available data and parameters for a
generic location $r$}\label{tab_data}
\begin{tabular*}{\textwidth}{@{\extracolsep{\fill}}lcccccccc@{}}
\hline
\multicolumn{2}{@{}l}{\textbf{Risk group}} & $\bolds{N}$ & $\bolds{\rho
}$ & $\bolds{\pi}$ & $\bolds{\delta}$ & $\bolds{\psi(\rho,\pi)}$ &
$\bolds{\psi(\pi,\delta)}$ & \multicolumn{1}{c@{}}{$\bolds{\psi(\rho,\pi
,\delta)}$} \\
\hline
Men & MSM & & $\checkmark$ & & & & $\checkmark$ & $\checkmark$ \\
& IDUs & & $\checkmark$ & $\checkmark$ & $\checkmark$ & & & $\checkmark
$ \\
& Born sub-Saharan Africa & & $\checkmark$ & & & & & $\checkmark$ \\
& STI clinic attendees & & $\checkmark$ & & & & $\checkmark$ &
$\checkmark$ \\
& Lower risk & & & & & & & \\
& ALL & $\checkmark$ & & & & & & $\checkmark$ \\[6pt]
Women & IDUs & & $\checkmark$ & $\checkmark$ & $\checkmark$ &
$\checkmark$ & & $\checkmark$ \\
& Born sub-Saharan Africa & & ${\checkmark}$ & ${\checkmark}$ &
$\checkmark$ & & & $\checkmark$ \\
& STI clinic attendees & & $\checkmark$ & & & $\checkmark$ &
$\checkmark$ & $\checkmark$ \\
& Lower risk & & & & & $\checkmark$ & & \\
& ALL & $\checkmark$ & & ${\checkmark}$ & & & & $\checkmark$ \\
\hline
\end{tabular*}
%
\end{table*}
%

\section{HIV Prevalence Estimation} \label{sec_prev}

Extending the notation introduced in Section~\ref{sec_intro}, HIV
prevalence $\pi_{t,r}$ in the general population at a single point in
time $t$ in location $r$ may be expressed as $\pi_{t,r} = \sum_g
\rho_{t,g,r}\pi_{t,g,r} = \sum_g \rho_{t,g,r}\pi_{t,g,r}\delta_{t,g,r}
+\break  \sum_g \rho_{t,g,r}\times \pi_{t,g,r}(1 - \delta_{t,g,r})$. The aim is to
estimate the basic parameters $\bolds{\theta}_{t,g,r} =
(\rho_{t,g,r}, \pi_{t,g,r}, \delta_{t,g,r})$. Having obtained the
posterior distribution of these, it is possible to obtain the
posterior distribution of any function of interest, {for example,} the
total number of infections $N_{t,r}\sum_g \rho_{t,g,r}\pi_{t,g,r}$ or
the total number of undiagnosed infections $N_{t,r}\sum_g
\rho_{t,g,r}\pi_{t,g,r}(1 - \delta_{t,g,r})$, where $N_{t,r}$ is the
location- and time-specific total population.
There are $13$ mutually exclusive risk groups defining a hierarchy of
risk. Men are {classified into the following}: men who have sex with
men [MSM
attending sexually transmitted infection (STI) clinics; MSM not
attending STI clinics; and past MSM]; injecting drug users (IDU,
current and past){;} heterosexual men born in sub-Saharan Africa (SSA);
heterosexual men attending STI clinics; and heterosexual men at low
risk (LR) of infection. Heterosexual women are {classified} in the same
way as heterosexual men. Geographically, there are three locations
(Inner London, Outer London, Rest of E\&W){,} and $t$ refers to the year
2008. In total there are $11\times3 + 13\times3+13\times3=111$
parameters as $\sum_g \rho_{t,g,r}=1$ for each gender.

\subsection{Data} \label{sec_data} Different types of data are
available on the following: group sizes, HIV prevalence, prevalence of
undiagnosed
infections, proportion of infections diagnosed, total number of
diagnosed infections and group distribution amongst diagnosed
cases. Data sources are described in full and commented upon elsewhere
(\cite{GoubarEtAl2008}, \cite{PresanisEtAl2010} and references therein), and
are only briefly reviewed here. Mid-year population estimates provide
information on $N_{t,r}$ and some risk group proportions
$\rho_{t,g,r}$. The remaining $\rho_{t,g,r}$ are derived from a
behavioural survey. Unlinked anonymous sero-prevalence surveys amongst
STI clinic attendees inform the prevalence of undiagnosed infection
$\pi_{t,g,r}(1 - \delta_{t,g,r})$. The analogous surveys amongst
pregnant women and IDUs inform prevalence $\pi_{t,g,r}$ and proportion
diagnosed $\delta_{t,g,r}$, both directly for some groups and
indirectly through functions of $\pi_{t,g,r},\delta_{t,g,r}$ and $\rho_{t,g,r}$. The
pregnant women's survey, in particular, measures prevalence in those
born in SSA and the remainder (NSSA). These NSSA are a mixture of STI
clinic attendees, IDUs{} and lower risk women. The observed data,
therefore, provide information on a complex function of HIV prevalence
in these groups {and} account for the probability of each group
being included in the sample. An annual cross-sectional survey of
diagnosed individuals collects information on functional parameters
representing both the total number living with diagnosed HIV
$ (N_{t,r}\sum_g \rho_{t,g,r}\pi_{t,g,r}\delta_{t,g,r} )$ and
the distribution of risk groups amongst these individuals $ (
(\rho_{t,g,r}\pi_{t,g,r}\delta_{t,g,r} )/ (\sum_g
\rho_{t,g,r}\pi_{t,g,r}\delta_{t,g,r} )  )$ for each
group $g$. Table~\ref{tab_data} summarises the spread and the type of
information available as well as the relationship between the
available data and the basic parameters, expressed here through
generic functions $\psi$. Note the sparseness of information on
heterosexual men and the multiplicity of data on heterosexual women.

\subsection{Inference} \label{sec_prevINFERENCE}

\subsubsection*{Sampling distributions}
Information $ \mathbf{y}_{t,g,r}$ from\break most sources can be expressed in
the form of count data $x_{t,g,r}$ with an associated denominator
$n_{t,g,r}$. These data can be {assumed to naturally} be realisations
of a {b}inomial random variable
\[
X_{t,g,r} \sim\operatorname{{b}inomial}(n_{t,g,r}, \psi_{t,g,r}),
\]
where $\psi_{t,g,r}$ equals any of $\rho_{t,g,r}$, $\pi_{t,g,r}$ and
$\delta_{t,g,r}$ if ${ \mathbf{y}}_{t,g,r}$ provides direct
information{} or is a function of these basic parameters.

The observed total numbers of diagnosed men and women in each
location, $x_{t,m,r}$ and $x_{t,f,r}$, respectively, are assumed to be
realisations of Poisson random variables $X_{t,m,r}\sim
\operatorname{Poisson}(\mu_{t,m,r})$ and $X_{t,f,r} \sim
\operatorname{Poisson}(\mu_{t,f,r})$, where
\begin{eqnarray*}
\mu_{t,m,r} & = & N_{t,m,r} \sum_{g_m}
(1 - \nu_{t,g_m})\delta _{t,g_m,r}\pi_{t,g_m,r}
\rho_{t,g_m,r},
\\
\mu_{t,f,r} & = & N_{t,f,r} \sum_{g_f}
(1 - \nu_{t,g_f})\delta _{t,g_f,r}\pi_{t,g_f,r}
\rho_{t,g_f,r}.
\end{eqnarray*}
Here $g_m$ and $g_f$ refer to male and female groups, respectively, and
$\nu_{t,g_m}, \nu_{t,g_f}$ are parameters representing potential bias
in the reported number of diagnosed individuals due to
nonattendance, under-reporting{} or duplication. The region-specific
numbers diagnosed in each risk group, $x_{t,g_m,r}$ and $x_{t,g_f,r}$,
are simultaneously drawn from gender-specific {m}ultinomial
distributions with size parameters $\mu_{t,m,r}$ and $\mu_{t,f,r}$,
and probability parameters
\begin{eqnarray*}
\xi_{t,g_m,r} & = & \bigl((1 - \nu_{t,g_m})\delta_{t,g_m,r}\pi
_{t,g_m,r}\rho_{t,g_m,r} \bigr)\\
&&{}\Big/\sum_{g_m}
(1 - \nu_{t,g_m})\delta _{t,g_m,r}\pi_{t,g_m,r}
\rho_{t,g_m,r},
\\
\xi_{t,g_f,r} & = & \bigl((1 - \nu_{t,g_f})\delta_{t,g_f,r}\pi
_{t,g_f,r}\rho_{t,g_f,r} \bigr)\\
&&{}\Big/\sum_{g_f}
(1 - \nu_{t,g_f})\delta _{t,g_f,r}\pi_{t,g_f,r}
\rho_{t,g_f,r}.
\end{eqnarray*}

The full likelihood $L_t(\bolds\theta_t;  \mathbf{y}_t)$
results from the product of each of these distributions, as
generically described in Section~\ref{sec_mpes}.

\subsubsection*{Sparseness of information}
One of the challenges to the ``direct'' methods is the lack of
information on some risk groups. Table~\ref{tab_data} clearly shows
that data on $\pi_{t,g,r}$ and $\delta_{t,g,r}$ for male heterosexuals
are sparse. This sparsity can be addressed by sharing information
between men and women. Although $\pi_{t,g,r}$ and $\delta_{t,g,r}$ are
expected to vary by gender and by location, it is reasonable to assume
that their male-to-female odds ratios might be similar between
regions. To borrow strength across locations and risk groups, the
following hierarchical structures are then assumed for the
male-to-female log odds ratios of prevalence $\mathrm{lor.}\pi_{t,g,r}$ and
proportion diagnosed $\mathrm{lor.}\delta_{t,g,r}$:
\begin{eqnarray*}
\operatorname{logit}(\pi_{t,g_m,r}) & = & \mathrm{lor.}\pi_{t,g,r} + \operatorname{logit}(
\pi_{t,g_f,r}),\\
\mathrm{lor.}\pi_{t,g,r}& \sim&\operatorname{{n}ormal}
\bigl(P_{t,g}, \sigma^2_{t,\pi}\bigr),
\\
\operatorname{logit}(\delta_{t,g_m,r}) & = & \mathrm{lor.}\delta_{t,g,r} + \operatorname{logit}(\delta
_{t,g_f,r}),\\
 \mathrm{lor.}\delta_{t,g,r} &\sim&\operatorname{{n}ormal}
\bigl(D_{t,g}, \sigma^2_{t,\delta}\bigr),
\end{eqnarray*}
%
%
with a further hierarchy over risk groups:
\[
P_{t,g} \sim\operatorname{{n}ormal}\bigl(\Pi_t,
\omega^2_{t,\pi}\bigr),\quad D_{t,g} \sim
\operatorname{{n}ormal}\bigl(\Delta_t, \omega^2_{t,\delta}
\bigr).
\]
%
The means $\Pi_t$ and $\Delta_t$ are {a priori} distributed as
{n}ormal$(0,100^2)$. The standard deviations
$\sigma_{t,\pi},\sigma_{t,\delta}$ and $\omega_{t,\delta}$ are given
informative priors expressing the belief that only $5\%$ of
region-specific male-to-female odds ratios (of both prevalence and
proportion diagnosed) will vary from the mean by {more than} a 1.3 factor
(Section~5.7.3 of \cite{SpiegelhalterEtAl2004}). The odds ratios for
prevalence are assumed to vary more across risk groups than across
regions, hence, the prior for $\omega_{t,\pi}$ is weaker: a~factor of
1.6 is used.

\begin{figure*}

\includegraphics{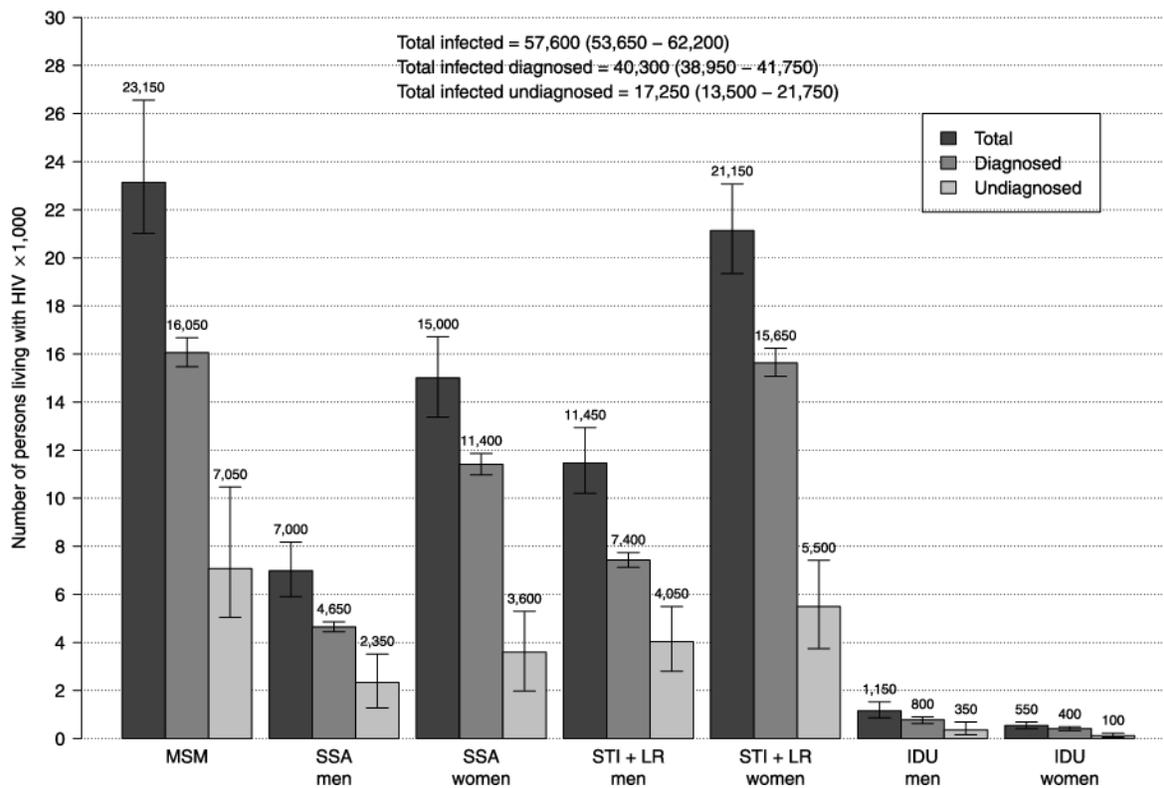}

\caption{Posterior median ($95\%$ credible
interval) number of HIV infections in adults aged 15--44 in E\&W in
2008, by diagnosis status and risk group.}\label{fig_prev2008}
\end{figure*}

\subsubsection*{Bias modelling and other indirect information}
A~further challenge to the estimation problem is represented by data
that indirectly inform a specific parameter of interest. The most
common example {occurs} where the data are known to be affected by biases,
as for the {above} total number of diagnosed infections. The parameters
$\nu_{t,g_m}$ and $\nu_{t,g_f}$ are, in this case, introduced to model
the known bias. In general, this is dealt with by introducing ``bias
models'' that take the generic form $\theta^{\prime} = \theta+ \varepsilon$ on
a suitable scale, where $\theta$ is the parameter of interest and
$\theta^{\prime}$ is the parameter directly informed by the data. The ``bias
parameter'' $\varepsilon$ is a measure of the discrepancy between
$\theta^{\prime}$ and $\theta$. Where information or expert opinion on the
size and/or direction of the bias is available, this is expressed as
an informative prior on~$\varepsilon$.\vadjust{\goodbreak}

Other challenges in the data sources, such as greater spatial
heterogeneity than is captured by the regional structure adopted in
the model, are met by more complex modelling, such as mixed effects
regression on a finer regional stratification. The parameters
$\psi_{t,g,r}$ in the {b}inomial expression above may therefore have a
more complex functional structure than the examples given here; see
\citet{GoubarEtAl2008}, \citet{PresanisEtAl2008},
\citet{PresanisThesis} for more
details.

\subsubsection*{Priors}
Diffuse {u}niform priors are chosen for the basic parameters
$\pi_{t,g,r}$ and $\delta_{t,g,r}$. The proportions of the male and
female populations in each risk group $\rho_{t,g_m,r}$ and
$\rho_{t,g_f,r}$ are given Dirichlet priors such that they sum to
$1$. Informative {n}ormal or {u}niform priors are assigned to bias
parameters such as $\nu_{t,g_m}$ and~$\nu_{t,g_f}$.

\subsubsection*{Results}
Samples from the posterior distribution are obtained using Markov
chain Monte Carlo (MCMC), implemented in \texttt{WinBUGS}
(\cite{winBUGS}). Posterior summaries are based on 8000 samples from
two chains after convergence is achieved.
Figure~\ref{fig_prev2008} presents the estimated number of HIV
infections in E\&W, by diagnosis status and risk group.

\section{Joint Prevalence and Incidence Model} \label{sec_joint}

Application of the prevalence model over successive years using a
sequence of data sets $\{\mathbf{y}_t\}, t \in1, \ldots, T$,
provides the joint posterior distribution of the proportions of the
population $N_{t,g,r}$ in each of three compartments: susceptible to
infection $s_{t,g,r} = \rho_{t,g,r}(1 - \pi_{t,g,r})$; HIV infected
but undiagnosed $u_{t,g,r} =  \rho_{t,g,r}\pi_{t,g,r}(1 -
\delta_{t,g,r})$; and infected and diagnosed $d_{t,g,r} =
\rho_{t,g,r}\pi_{t,g,r}\delta_{t,g,r}$. These can be interpreted as
estimates of the state at time $t$ of a dynamical system describing
the processes of infection and diagnosis. Such a system can be
approximated by a continuous-time Markov model whose dynamics are
described through a system of ordinary differential equations
(ODEs). As in \citet{PresanisEtAl2011}, we focus here on the MSM group
and, for simplicity, drop the subscripts $g$ and $r$.
\setcounter{figure}{2}
\begin{figure}

\includegraphics{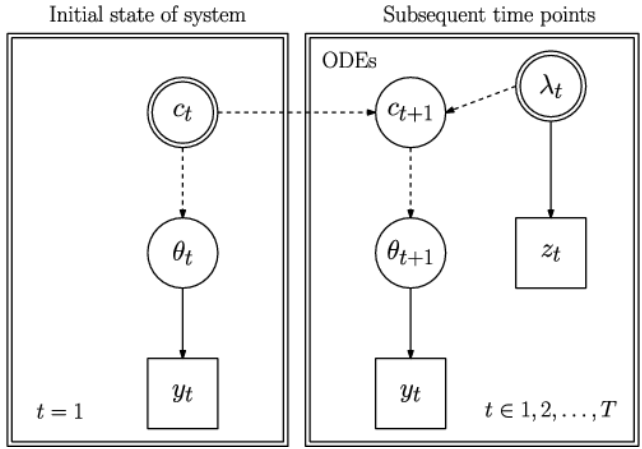}

\caption{Schematic DAG of the joint prevalence and
incidence model.}\label{fig_joint}\vspace*{6pt}
\end{figure}
Let ${\mathbf{c}}_t = (s_t,u_t,d_t)$ and denote by
$\bolds\lambda_t$ the transition rates in the time interval
$[t,t+1)$, assumed piecewise constant over the interval. Using
additional data $\{{\mathbf{z}}_t\}$ (\cite{PresanisEtAl2011}) and
prior information on demographics and risk behaviour uptake, a joint
prevalence and incidence model can be formulated to allow simultaneous
estimation of the prevalence parameters $\bolds{\theta}_t$, the
compartment proportions ${\mathbf{c}}_t$ and the transition rates
$\bolds{\lambda}_t$ including $\lambda^{s,u}_t$, the incidence
rate, that is, the rate at which susceptible individuals enter the
infected state. The DAG in Figure~\ref{fig_joint} provides a schematic
representation of this joint model. The proportions
${\mathbf{c}}_{t+1}$ at time $t+1$ are defined, through the ODEs, in
terms of the rates $\bolds{\lambda}_t$ during the period
$[t,t+1)$ and the initial condition of the system at $t=1$. The
prevalence parameters $\bolds{\theta}_t$ and
$\bolds{\lambda}_t$ govern the prevalence and rate data,
respectively. Note that this DAG has the same structure as that in
Figure~\ref{fig DAG}. Now the $\bolds{\lambda}_t$ and
${\mathbf{c}}_1$ are the \textit{basic} parameters and the
$\bolds{\theta}_t$ are functional parameters.\looseness=1

Inference is conducted as described in Section~\ref{sec_mpes}. The
likelihood of the joint data
is
\[
L({\mathbf{c}}_1,\bolds{\lambda}; {\mathbf{y}}, {\mathbf{z}})=\prod
_{t=1}^{T}L_t({
\mathbf{c}}_1; {\mathbf{y}}_t)L_t(\bolds{
\lambda}_t;{\mathbf{z}}_t),
\]
where
$L_t(\bolds{\lambda}_t;{\mathbf{z}}_t$) is the likelihood
contribution of the demographic and behavioural data informing
transition rates. Assuming independent vague priors for
$\bolds{\lambda}_t$ and a $\operatorname{Dirichlet}(1,1,1,1)$ prior for the
compartment proportions at $t=1$, ${\mathbf{c}}_1$, the joint
posterior distribution for ${\mathbf{c}}_1$ and
$\bolds{\lambda}$, and therefore also of $\bolds{\theta}$, is
obtained through MCMC implemented in WinBUGS. Note that the likelihood
contribution of the prevalence data, ${\mathbf{y}_t}$, depends on
the ${\mathbf{c}_{t}}$, the ODEs' solutions, which are derived
numerically for the current parameter values at each MCMC iteration
using the Runge-Kutta algorithm in the \texttt{WBDiff} package in
\texttt{WinBUGS}.
Figure~\ref{fig_jointPost} shows posterior distributions resulting
from the joint prevalence and incidence model.
\setcounter{figure}{3}
\begin{figure*}

\includegraphics{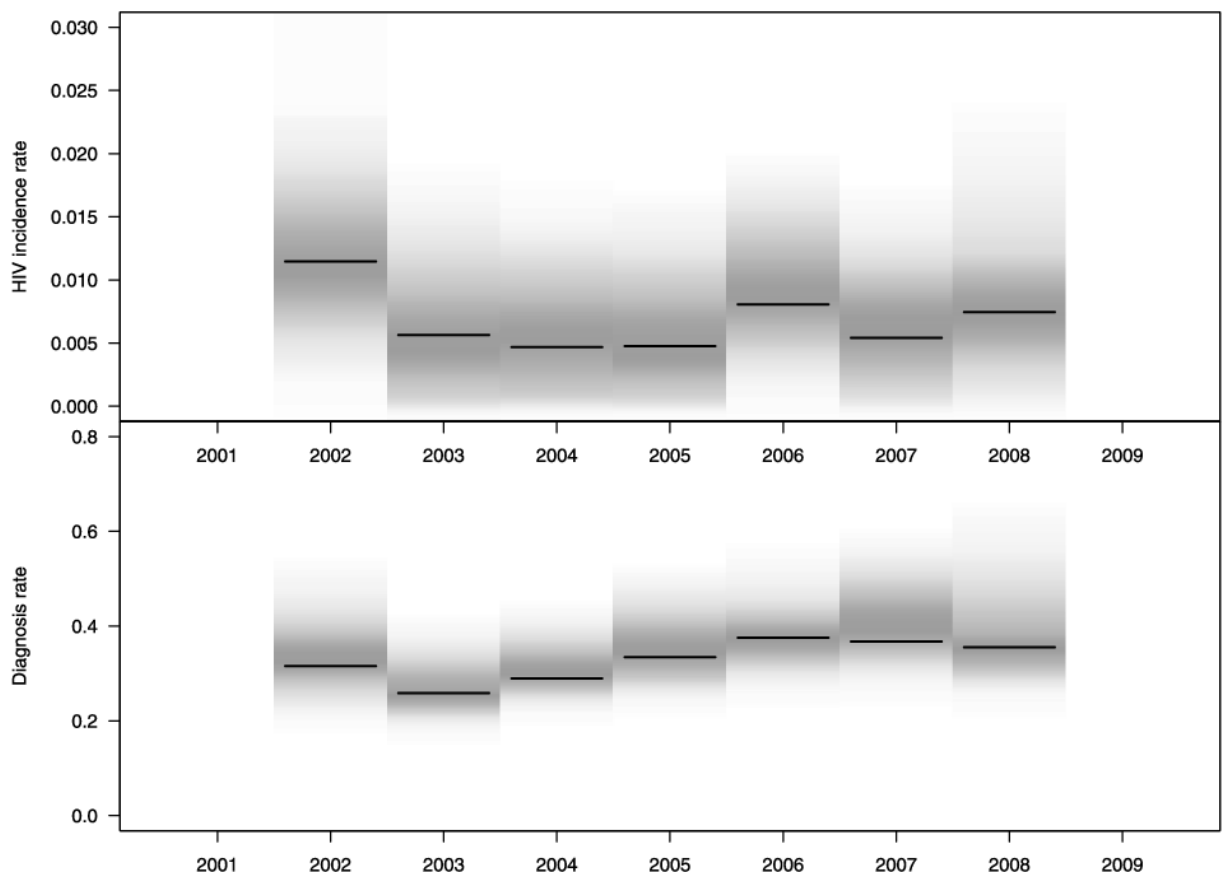}

\caption{Density strip plots of posterior distributions of incidence and
diagnosis rates in MSM (darker colour corresponds to higher density and
horizontal black line denotes the posterior median), 2002--2008.}\label{fig_jointPost}
\end{figure*}

\section{Discussion} \label{sec_discuss}

From a methodological point of view, this work has responded formally
to the need, perceived by epidemiologists working in the HIV arena, to
``triangulate'' \textit{all} information: from multiple and imperfect
sources and expert opinion on the epidemiological interpretation of
the data from these sources. The approach is clearly appealing: it
uses data fully, minimising potential biases due to selection of
information; it typically leads to more precise estimates, which are
consistent with all information; {and} it accounts for
all sources of uncertainty, naturally reflected in the posterior
distributions of parameters and quantities of interest.

\subsection{Why Bayesian?}

In principle, evidence synthesis does not need to be carried out in a
Bayesian framework; see, {for instance,} \citet{EddyEtAl1992} and much of
the meta-analytical work referenced in \citet{sutabr:00}. Indeed, work
exists to estimate HIV prevalence in a single risk group by
synthesising three data sources in a classical approach, accounting
for missing data (\cite{WalkerEtAl2011}). However, the unprecedented
multiplicity of data sources, risk groups{} and indirect information
involved in the work described here requires a Bayesian approach, with
clear benefits over classical, likelihood-based alternatives. The main
{advantage} is the ability to (i) explicitly introduce and
(ii) formally quantify expert judgements. The hierarchical
model introduced in Section~\ref{sec_prevINFERENCE} to tackle data
sparseness offers such an example: only through reasonably chosen
informative priors on the standard deviation hyperparameters has it
been possible to overcome identifiability problems due to lack of
information (see sensitivity analyses in \cite{PresanisThesis}).
Second, a~Bayesian model can be easily extended to include auxiliary
``bias'' parameters to quantify lack of validity and relevance of data
items for the estimation {for} any specific parameter. Expert
epidemiological information on the direction and magnitude of such
biases is naturally accommodated in a Bayesian setup through
carefully chosen priors (see Section~\ref{sec_prevINFERENCE} and
references therein). It is not immediately obvious how a classical
modelling approach would accommodate such information. Computational
convenience represents a further advantage of a Bayesian approach. As
the posterior distribution is estimated through simulation, it is
straightforward to obtain inferences on any functional parameter of
interest. The likelihood function of even a moderately sized evidence
network is unlikely to be sufficiently tractable to allow comparably
streamlined inference.

\subsection{Impact on the Real World} Since 2005 our
``multi-parameter'' evidence synthesis has been the approach adopted to
produce the official estimates of the magnitude of the HIV problem in
the UK, in particular, the undiagnosed component, underlying current
testing recommendations
(\href{http://www.hpa.org.uk/web/HPAweb&Page&HPAwebAutoListName/Page/1201094588821}{http://www.hpa.org.uk/web/HPAweb\&Page\&}
\href{http://www.hpa.org.uk/web/HPAweb&Page&HPAwebAutoListName/Page/1201094588821}{HPAwebAutoListName/Page/1201094588821}).
In 2011, estimated trends on the prevalence of undiagnosed infection and
incidence in MSM informed the work of the House of Lords Select
Committee on HIV/AIDS in the UK
(\href{http://www.publications.parliament.uk/pa/ld201012/ldselect/ldaids/188/188.pdf}{http://www.publications.}
\href{http://www.publications.parliament.uk/pa/ld201012/ldselect/ldaids/188/188.pdf}{parliament.uk/pa/ld201012/ldselect/ldaids/188/188.}
\href{http://www.publications.parliament.uk/pa/ld201012/ldselect/ldaids/188/188.pdf}{pdf}).
Recently, dissemination
of the method has also attracted the interest of international public
health organisations. Funded by the World Health Organization as part
of an ongoing critical review of current methods for HIV prevalence
estimation in concentrated epidemics (\cite{GhysEtAl2008}), the
prevalence model has been adapted to estimate HIV burden in the
Netherlands {for the year 2007} (\cite{ContiEtAl2011}). In comparison
to other direct-type
methods, the evidence synthesis approach was found to be the most
flexible and statistically sound (\cite{vanVeenEtAl2011}).

\subsection{Current Challenges}

The model building and criticism processes in this work have led to a
critical understanding of the\break strengths and weaknesses of the various
sources of HIV information available in the UK, often challenging
common interpretation of the data. Extensive sensitivity analyses
have been carried out for prior and structural
assumptions, to the sampling distributions employed, as well as to the
data sources included
(\cite{GoubarEtAl2008}, \citeauthor{PresanisEtAl2008} \citeyear{PresanisEtAl2008}, \citeyear{PresanisEtAl2011}, \cite{PresanisThesis}). Moreover,
routine {annual} application of the model has led to
continual model development, responding to changes in surveillance,
the availability of new data sources{,} and ongoing
model criticism in the cycle recommended by \citet{Box1980} and
\citet{OHagan2003}, amongst others. Some of the development required
and in progress includes addressing issues of missing data, using
ideas as in \citet{WalkerEtAl2011} and a comprehensive model of the
process of diagnosis in STI clinics based on a new surveillance system
(\href{http://www.hpa.org.uk/Topics/InfectiousDiseases/InfectionsAZ/HIV/OverallHIVPrevalence/}{http://www.hpa.org.}
\href{http://www.hpa.org.uk/Topics/InfectiousDiseases/InfectionsAZ/HIV/OverallHIVPrevalence/}{uk/Topics/InfectiousDiseases/InfectionsAZ/HIV/}\break
\href{http://www.hpa.org.uk/Topics/InfectiousDiseases/InfectionsAZ/HIV/OverallHIVPrevalence/}{OverallHIVPrevalence/}).

More generally, model criticism is essential in an evidence synthesis
approach. As data come from multiple sources depending on shared
parameters, it becomes crucial to understand and communicate which
sources (including priors) drive conclusions and whether the various
items of evidence result in consistent or conflicting
inference. Efforts clearly need to be focussed on the development of
transparent methods for model assessment and criticism, given that
evidence synthesis approaches are being increasingly employed in
different areas of science.

In the same spirit, an important step toward improved communication
and dissemination of
Bayes\-ian evidence synthesis would be the availability of a
user-friendly computing environment, facilitating access and
implementation of the methodology to
nonexperts. \citet{vanVeenEtAl2011} also identified the lack of such a modelling
interface as a restriction to the more widespread adoption of our approach.

\section*{Acknowledgements}
We thank the HIV division at the Health Protection
Agency for providing the data on which this work is based. Special
thanks go to Dr{.}~Valerie Delpech and
Prof{essor}~Noel Gill for discussion on the
interpretation of data and results. We are indebted also to
Prof{essor}~David Spiegelhalter for many useful
methodological discussions over the last few years. This work was
supported by the Medical Research Council [grant number G0600675,
Unit Programme Number U105260566] and the Public Health England.
%


\end{document}